%% file: templateArxiv.tex
\documentclass{article}
\usepackage{amsmath,amssymb,amsfonts}
\usepackage[a4paper]{geometry}
\usepackage{caption}
\usepackage{pifont}
\usepackage{eqparbox}
\usepackage{url}
\usepackage{multirow}
\usepackage{textcomp}
\usepackage{bm}
\usepackage{dblfloatfix}
\usepackage{graphicx}
\usepackage{anyfontsize}
\usepackage[square,numbers,sort&compress]{natbib}
\usepackage[colorlinks=true, allcolors=blue, urlcolor=black]{hyperref}

\newgeometry{
  left=2.2cm,
  right=2.2cm
}

\usepackage[affil-it]{authblk}

\title{Hybrid Analytical--EMT Method for HVDC Protection System Component-Level Design}
\author[1,2]{Abolfazl Mohammadi\thanks{This work is part of the DIRECTIONS project, supported by the Energy Transition Fund, DG Economy, Belgium and partially supported by the Energy Transmission Competence Hub (Etch). 
The authors are with Energy Transmission Competence Hub
(Etch), EnergyVille, 3600 Genk, Belgium, and also with Electrical
Energy Systems and Applications (ELECTA), Department of Electrical Engineering (ESAT), KU Leuven, 3001 Leuven, Belgium (e-mail:
{abolfazl.mohammadi@kuleuven.be}).}}
\author[1,2]{Merijn Van Deyck}
\author[1,2]{Geraint Chaffey}
\author[1,2]{Dirk Van Hertem}
\affil[1]{Department of Electrical Engineering, KU Leuven, Belgium}
\affil[2]{Etch-EnergyVille, Belgium}
\date{}

\begin{document}
\maketitle

\input{0-Abstarct}

\section*{Keywords}
HVDC protection, DC circuit breaker , DC inductor sizing, EMT modeling.

\input{1-Introduction}
\input{2-Method}
\input{3-Case_Studies}

\input{4-Results}
\input{5-Conclusion}

\bibliography{References}
\bibliographystyle{IEEEtran}

\end{document}

%% file: 0-Abstarct.tex
\begin{abstract} 

Protection system design for multi-terminal HVDC grids is challenging due to the complexity of the system and the often conflicting design requirements. Effective specification of protection component parameters (e.g., DC circuit breakers and series DC inductors) during component-level design is crucial due to interdependencies among components, the need for detailed modeling, and the complex interactions between the protection system and converter control systems. Both analytical and simulation-based approaches have been proposed as solutions for component-level design. However, analytical methods may not accurately represent system behavior given that approximation is necessary, and simulation-based approaches often require extensive computational effort and time. Therefore, this paper presents an efficient systematic design method, combining both approaches. First, a fundamental analytical solution is derived to consider the protection system requirements. Then, a hybrid analytical–EMT methodology is proposed to accelerate convergence toward the required design parameters, after which detailed models are applied to ensure accuracy in design and validation. The approach is applicable to component-level design for both fully and partially selective protection strategies in HVDC grids.

\end{abstract}

%% file: 1-Introduction.tex
\section{Introduction}
\label{1-introduction}

{M}{ulti--terminal} HVDC systems provide a promising solution for future transmission networks, enabling bulk power transfer over long distances, interconnection of asynchronous AC systems, and integration of offshore wind farms~\cite{Dirk2010DCgrids}. Therefore, the industry is increasingly looking toward large-scale HVDC grids to transmit energy across Europe and to achieve the required expansion in electricity transmission capacity~\cite{ENTSO-E}. However, one of the main technical challenges in HVDC grids is protection against DC faults. Due to the characteristics of DC fault currents, such as the absence of natural current zero crossing and a rapid rise rate, dedicated protection devices such as a DC circuit breaker (DCCB) for interrupting fault currents, and a DC inductor for limiting the fault current may be be needed for multi-terminal HVDC systems~\cite{Dirk2010DCgrids,SysDCB2020Mudar}. Protection systems that allow DC-side fault isolation help avoiding prolonged interruptions of power transfer, ensuring continuity of supply, and maintaining the operation of the healthy parts of the grid~\cite{promotion2017d42}. 

System-level protection design, which includes the specification of the system topology, protection strategy and converter DC fault ride-through (DC-FRT) capabilities~\cite{Trondheim,northsea2023}, provides inputs for the component-level design. In this component-level design stage, key protection system parameters, such as DCCB characteristics and fault current limiting inductance, are specified to meet the system-level and component-level requirements~\cite{Trondheim, Rector2018,SysDCB2020Mudar}. 
Component-level design for multi-terminal grids is not straightforward, as components, such as DCCBs, DC inductors, and converter control systems, are interdependent~\cite{Patrick2023}. Therefore, detailed electromagnetic transient (EMT) modeling is typically required for these studies. In industrial applications such as the Zhangbei project~\cite{Zhangbei-2018}, DC inductors installed in series with DCCBs are reported in the range of 150--200~mH to limit the DC fault current. Other studies propose inductances exceeding 500~mH for this purpose~\cite{interopera2025D21,Patrick2023}. Large inductance values increase installation costs~\cite{kontos2016reactor} and may lead to stability issues~\cite{elsodany2026damping,Wang-2016}. Therefore, DC protection systems should be properly designed to account for these effects by sizing DC inductors as small as possible while still satisfying protection requirements.

Existing design methods take different approaches, which can be categorized into three main groups. The first group is based on analytical solutions, such as fault-current estimation~\cite{Tang2026,DCfault,Leu2022,mudar2022dc}. Such methods have been used for DC inductor sizing~\cite{Rector2018} and DCCB parameter specification~\cite{SysDCB2020Mudar}. However, analytical methods cannot suitably represent the complex behavior of the HVDC system~\cite{abedrabbo2021continuous}. In practice, representing components such as HVDC cables or converters fully analytically is impractical and overly complex~\cite{abedrabbo2021continuous}, as details are often unavailable due to IP concerns and black-box representations~\cite{Luscan}. The second type of design method, such as~\cite{DCR-ACDC2026,hart2025impact}, is based on exhaustive EMT  simulations. Although these methods can be more representative of system and component behavior, in practice, relying solely on EMT simulations to design components often requires extensive, repetitive simulations, with hundreds to thousands of iterations, depending on the system size. The third category uses hybrid methods, a combination of methods and tools, to tackle the problem. For instance, in~\cite{Ilka2021}, optimization-based methods are used to size the DC inductor optimally. These methods embed an EMT model into an optimization problem to optimally size components and represent system behavior more accurately. However, these optimization problems require  expensive EMT simulations, particularly as the system size increases towards future large-scale multi-terminal HVDC systems.

\begin{figure}[t]
    \centering
    \includegraphics[clip, trim=0.6cm 0cm 0.6cm 0cm, width=0.6\linewidth]{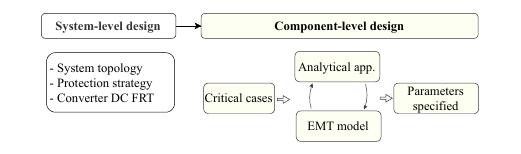}
    \caption{Overview of the protection component sizing method}
    \label{fig:Overview}
\end{figure}

In this paper, an efficient hybrid method for designing protection components is proposed. An overview of the proposed method is illustrated in Fig.~\ref{fig:Overview}. The method first identifies critical design scenarios to reduce the number of simulation cases. Then, a hybrid methodology is applied, where combination of analytical approximations and EMT model allowing accurate specification of the component parameters. The main contributions of this paper are as follows:
\begin{itemize} 
    \item Development of a computationally efficient, systematic method for sizing protection components for both fully and partially selective DC protection systems.
    \item Identification of critical cases to reduce the number of simulations for the design, increasing the efficiency of the design process.
    \item Achieving high-accuracy component parameter specification using the proposed hybrid analytical–EMT method, with the possibility of incorporating vendor models.

\end{itemize}
The rest of the paper is structured as follows. Section II outlines the protection systems requirements, followed by the component-level design method in Section III. Section IV describes two case studies, and Section V discusses the results.

%% file: 2-Method.tex
\section{Protection System Functional Requirements}

\subsection{System-level requirements}

System-level functional requirements of HVDC protection systems have recently been developed~\cite{promotion2017d42,Leterme2019,Trondheim,SysDCB2020Mudar}. Key system-level design aspects include the loss-of-infeed limits of the surrounding AC grids~\cite{promotion2017d42}, the selection of protection strategies and zone configurations (i.e., fully selective, partially selective, or non-selective)~\cite{Leterme2019,Trondheim}, and the converter DC-FRT requirements~\cite{promotion2017d42,SysDCB2020Mudar}. These requirements should be specified during the system-level protection design stage to ensure that high-level objectives, such as the stability of the surrounding AC and DC grids, are maintained~\cite{Trondheim}. Therefore, in this paper as shown in Fig.\ref{fig:Overview}, the fundamental inputs requirements form system-level design for component-level design are as follows:
\begin{itemize}
    \item System topology and configuration
    \item Protection strategy and protection zones boundaries
    \item Converter DC-FRT requirements 
\end{itemize}

\subsection{Component-level requirements}
The design of the selected components must ensure that the protection system can effectively isolate faults while satisfying the individual component constraints, such as DCCB and converter maximum current capacity~\cite{Leterme2019}. While the system-level design provides the system-level constraints, there is flexibility in the sizing of the individual components that make up the protection system, to achieve the same system-level requirements. In this paper, component-level design is considered to be the sizing or specification of any individual components. This paper focuses primarily on components that are co-dependent and require co-design~\cite{Trondheim}. These include DC inductor size, converter overcurrent capacity, and DCCB parameters such as operation time and DC fault neutralization time (sum of DCCB internal commutation time and protection relay time)~\cite{interopera2025D21,Rector2018,SysDCB2020Mudar}. Other protection related components, such as protection algorithm constraints, instrument transformers, and other primary equipment, may not necessarily require co-design, and are hence not considered in the examples in this paper -- although the method could be extended to include other components where deemed relevant.

%---------------------------------------------------------------------------------
\section{Deriving Component-Level Design Constraints}
This section presents the derivation of the necessary analytical approximations to represent the component-level functional requirements and constraints used in the design method.  

To meet the protection component-level design objectives, several variables can be adjusted. DCCB operation time, DC reactor size, and converter overcurrent capacity are all co-dependent and should each be considered in the design stage. For example, assuming the converter's overcurrent capability is fixed, a larger inductor or a faster circuit breaker can be used to meet all design objectives. Although there are other novel fault-current handling methods, such as~\cite{Ldccontrol2020,Cwikowski-2018}, this paper relies on more industrially relevant technology.  

The DCCB operation time~\cite{DCCB2020China,scibreak2020} and converter overcurrent capacity~\cite{Patrick2023} are generally limited to a certain range due to technical constraints in industrial technology. Hence, the DC inductor size is considered in this paper as the primary design variable  First, the component-sizing method is developed for a single protection zone, and then a more general approach is derived. 

\subsection{Converters functional requirements and constraints}

For converter DC FRT requirements, according to~\cite{IECTS}, converters in the healthy zone may be required to remain operational (continuous operation), temporarily block, or permanently block following a DC fault. This paper primarily considers the first option, in which no converter is allowed to block in the healthy zone, as this imposes more stringent conditions on the converter and DC inductor~\cite{SysDCB2020Mudar}. Otherwise, converters will not impose requirements on DC indoor sizing. 

\begin{figure}
     \centering
     \includegraphics[clip, trim=0.3cm 0.3cm 0.3cm 0.5cm,width=0.6\linewidth]{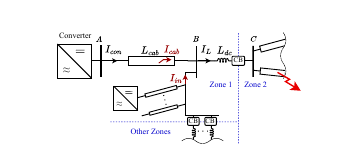}
      
     \caption{A HVDC grid for component sizing for a single zone}
     \label{fig:sys_for_derive_Lcon}
 \end{figure}

As an input from the system-level design, it is assumed that the system topology with DCCBs is known. A generic HVDC grid is shown in Fig.~\ref{fig:sys_for_derive_Lcon}, where the grid is divided into multiple protection zones by DCCBs. For the design of a single protection zone, the method focuses on one zone (Zone~1), while the other zones are represented in an aggregated manner. Although having only a converter in Zone~1 and directly connected to the DCCB can be considered as a worst case~\cite{ACDC2026} to say connected during the fault, adding additional converters and cables within Zone~1 makes the method sufficiently general to represent both partially and fully selective strategies. Hereafter, the term `zone' refers to a protection zone. Furthermore, the inductor in the system is assumed to be located in series with the DCCB, as shown in Fig.~\ref{fig:sys_for_derive_Lcon}. 

To size the inductor based on converter limits, the relationship between the DC inductor current and the converter current must be approximated. Following a fault in Zone~2, the cables and converters in other zones feed current into the fault, through the DC inductor $L_{dc}$. Therefore, the DC inductor current can be approximated as $I_{L} = I_{con}+I_{cab}+I_{in}$, which consists of converter current $I_{con}$, connection cable capacitive discharge current $I_{cab}$, and the infeed current from the adjacent cables and zones $I_{in}$. The represented aggregated infeed current ($I_{in}$) and connection cable capacitive discharge current ($I_{cab}$), rather than relying on a simplified analytical solution, can be measured from a detailed EMT model.

By applying a fault in Zone~2 along the cable, the voltage at bus~B can be approximated using~(\ref{eq:UB1}) for the time window of interest, i.e., the neutralization time ($t_n$). This time is the sum of the protection relay time and the DCCB internal current commutation time ($t_n = t_{relay} + t_{cb}$)~\cite{CIGRE-TB683}.  
\begin{subequations} \label{eq:UB1}
\begin{align}
U_{B} &\approx L_{dc} \frac{\Delta I_{L}}{t_n} + U_{C} \\
\Delta I_{L} &\approx \Delta I_{in} + \Delta I_{cab} + \Delta I_{con}
\end{align}
\end{subequations}

In this equation, $\Delta I_{L}$ is the change in the inductor current over the interval $t_n$, where $\Delta I_{in}$ and $\Delta I_{cab}$ represents the change of infeed and connection cable current changes, and $\Delta I_{con}$ is the converter DC side current change over the given period. $U_{C}$ is the cable-end voltage on Bus C, which highly depends on the cable parameters, the inductor value, and the fault location. The cable voltage envelope can be estimated by simulating various fault locations along the DC cable and taking the minimum of average voltage over the $t_n$ as $\bar{U}_{m} = \min(\bar{U}_{C}, 0)$, where $\bar{U}_C$ is average of $U_C$ over $t_n$ \cite{interopera2025D21}.

Considering Fig.~\ref{fig:sys_for_derive_Lcon}, $U_B$ can also be expressed as~(\ref{eq:UB2}) by applying KVL from the converter location to bus~B. The converter resistance and the connection cable resistance are assumed to have a negligible impact. To maintain the continuous operation of the converter, its control should remain active during fault neutralization. Therefore, in (\ref{eq:UB2}), it is assumed that the converter behaves as a voltage source and its terminal voltage is  $U_{dc}$, equal to the nominal value~\cite{Patrick2023}. In \eqref{eq:UB2},  $L_{eq}=L_{con}+L_{cab} $ and $L_{con}=(2/3)L_{arm}$ is the converter equivalent inductor from terminal and $L_{cab} $ is the lumped inductance of the connection cable.
\begin{equation} \label{eq:UB2}
U_{B} \approx U_{dc} - L_{eq} \frac{\Delta I_{con} }{t_n}
\end{equation}

Based on (\ref{eq:UB1}) and (\ref{eq:UB2}), the DC inductor for converter requirement, $L_{dc}^{con}$, can be expressed as:
\begin{equation} \label{eq:L_con}
    L_{d c}^{con} \approx 
    \frac {(U_{dc}-\bar{U}_{m}) t_n-
    L_{eq}  \Delta I_{con} }
    {\Delta I_{con} + \Delta I_{in} + \Delta I_{cab} }  
\end{equation}

Considering \eqref{eq:L_con}, $\Delta I_{in}$ and $\Delta I_{cab}$ are measurements that come from the EMT model used to approximate the inductor value, while the other parameters are grid parameters that are assumed to be known. To size the minimum inductor to meet the converter requirements, $\Delta I_{con}$ needs to be set to its critical value. The worst-case set point corresponds to the converter operating at its rated current in rectifier mode ($I_{con}^r$) with maximum reactive power, leading to the critical value of $\Delta I_{con}^c = I_{con}^m - I_{con}^{r}$. The converter DC terminal current limit ($I_{con}^{m}$) can be derived based on its arm current limit as \eqref{eq:I_con_max}~\cite{Patrick2023}, where $K$ is the converter overcurrent factor or per-unit blocking threshold, $i_a^r$ is the rated peak arm current, and $i_{ac}^{max}$ is the maximum AC current limit on the converter side. Equation \eqref{eq:I_con_max} is modified to use $i_{ac}^{max}$ instead of the rated AC current to account for the actual AC current contribution. Here, $i_{ac}^{max}$ is a limit imposed by the converter current controller~\cite{TB604}.  
\begin{equation} \label{eq:I_con_max}
  I_{con}^{m} = 3\,\left(K\,i_{a}^{r}-\frac{i_{ac}^{max}}{2} \right) 
\end{equation}

Based on \eqref{eq:L_con}, parameters such as overcurrent capacity, maximum AC current limit, protection relay delay, and DCCB technology affect the critical inductor value. Furthermore, it is evident from (\ref{eq:L_con}) that the system configuration affects the required inductor value. For example, a lack of adjacent cables ($\Delta I_{in}$) and connection cable ($\Delta I_{cab}$) results in a large value for $L_{dc}^{con}$. Overall, the converter requirements may impose a minimum inductance requirement for the DC inductor. 
 
\subsection{DCCB requirements and constraints}
As the DCCB current should not exceed its maximum limit, the inductor in series with the DCCB must be designed to meet this constraint. According to \eqref{eq:UB1}, considering the maximum voltage difference across the inductor and substituting the critical value of $\Delta I_L$ provides an approximation of the critical inductor value. For the maximum voltage difference $\Delta U = U_{B}-\bar{U}_m$, $U_{B}$ is assumed constant and equal to the nominal system voltage, and $\bar{U}_m$ is the cable voltage envelope that accounts for the traveling wave effect. The critical value for the current change is $\Delta I_L^c = I_{cb}^m-I_{cb}^r$, where $I_{cb}^m$ and $I_{cb}^r$ are the DCCB maximum current limit and rated current, respectively. As a result, the critical inductor size based on the DCCB requirements is approximated by \eqref{eq:Lcb}, where a lower maximum current limit, a longer operating time, and a higher rated current result in more stringent requirements for the inductor value. 
\begin{equation} \label{eq:Lcb}
    L_{dc}^{cb} \approx \frac{U_{dc}-\bar{U}_{m}}{I_{cb}^m- I_{cb}^{r}}t_n
\end{equation}

\begin{figure}
    \centering
    \includegraphics[clip, trim=0.4cm 0cm 0cm 0.2cm,width=0.4\linewidth]{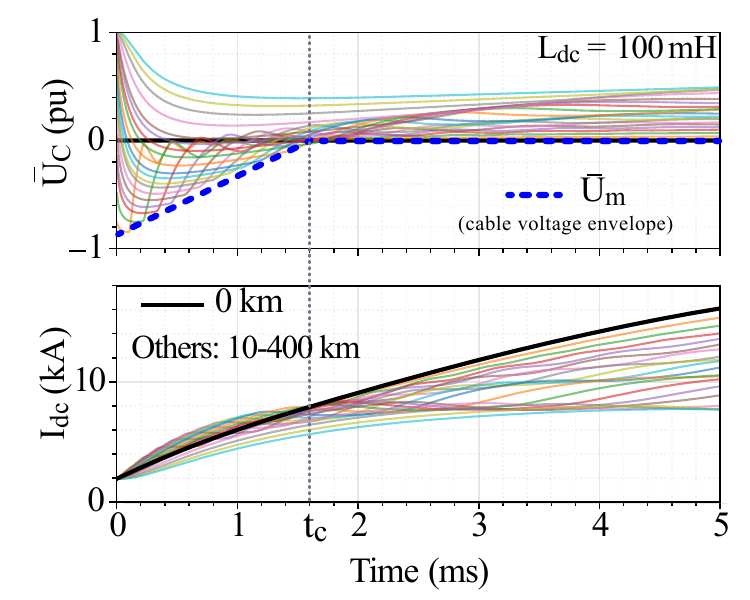}
    
   \caption{Critical fault location along the cable according to cable voltage envelope ($\bar{U}_m $)}
   \label{fig:CFL}
\end{figure}

\subsection{Critical cases for design}
\label{subsec:critical cases}
For component-level design, it is essential to identify the critical operating conditions of components and base the design on them. This approach ensures that the selected design parameters remain valid across all relevant scenarios while minimizing the total number of cases that need to be evaluated.
\subsubsection{Critical converter} Within each zone, a critical converter that reaches its overcurrent limit faster than the others for out-of-zone faults must be identified. According to equation~(\ref{eq:L_con}), the critical converter in a zone depends on both system configuration, such as line parameters, and also converter parameters as shown in \eqref{eq:I_con_max}. 
In equation~(\ref{eq:L_con}), the assumption of zero infeed current ($\Delta I_{in} = 0$) and zero cable discharge current ($\Delta I_{cab} = 0$) corresponds to a worst-case condition for the converter. Accordingly, this case requires the largest inductor to prevent blocking within the fault neutralization time. Therefore, the critical converter in each zone is identified as the one requiring the largest inductor. For instance, if two converters in a protection zone have the same parameters, the one connected directly to the DCCB or via the shortest line is the critical one due to the low fault wave propagation delay and low $\Delta I_{cab}$ ~\cite{ACDC2026}. Once the critical converter for each zone is identified, the corresponding infeed current can be determined in the system.

\subsubsection{Critical power flows} The identification of critical power flows in the system is relatively straightforward. The power flows that result in the maximum loading of the critical converter in rectifier mode and the maximum loading for the DCCB under study, are considered critical power flows with regard to a DC-side fault~\cite{SysDCB2020Mudar}.

\subsubsection{Critical fault type and location} The critical fault type depends on system configuration and grounding options, e.g., a pole-to-ground (P--g) fault in a bipolar system is a possible fault for a cable-based system with zero fault resistance~\cite{Leterme2019,Ground2014}. The worst-case fault location is the one that results in a rapid rise and high peak fault current. Due to the traveling-wave effect, the reflection coefficient, the cable parameters, and fault neutralization time, it is possible that a critical fault location lies along the cable in the other zone~\cite{cwikowski2016}.
As shown in Fig.~\ref{fig:CFL}, the critical fault location depends on the fault neutralization time ($t_n$) and the critical time ($t_c$) where $\bar{U}_m \leq 0$. For $t_n \ge t_c$, a terminal fault (0~km) results in a minimum voltage envelope ($\bar{U}_{m} = 0$) and a high fault current after the critical time (e.g., $t_c = 1.8$~ms in the illustrated case with example cable parameters). However, if $t_n < t_c$, a non-terminal fault should be considered as the critical fault location. This critical time ($t_c$) depends on parameters such as cable characteristics and the DC inductor, and it must be determined for each system using a similar approach as shown in Fig.~\ref{fig:CFL}. It is worth noting that although a large DC inductor can result in smaller $\bar{U}_{m}$ and increase the  $t_c$, it more limits the fault current.

%-----------------------------------------------------------------------------------------
\section{Component-level design algorithm}

After identifying the system-level and component-level requirements of the protection system, the protection component can be designed and specified to satisfy them. This section applies a hybrid analytical and EMT simulation-based approach.

\subsection{Core component sizing algorithm}
The key design parameters for the system are the converter overcurrent capability, the DCCB current-interruption capability, the operation time, and the DC inductor value. Due to the interdependence among these parameters~\cite{interopera2025D21}, the inductor value is treated as the output in this paper. The others are treated as inputs to the algorithm, as their ranges are limited by current DCCB and converter technologies. The core algorithm for single DC inductor sizing is shown in Fig.~\ref{fig:core-algorithm} and described below.  

\subsubsection{Part A -- System-level requirements and critical cases identification}

As a starting point, the system topology must be known, with protection zones and their boundaries primarily defined by the DCCB locations. The DC-FRT requirement for each converter is specified in a system-level study. In this paper, it is assumed that converters in the healthy zone are not allowed to block. Otherwise, from the converter perspective, there will be no hard constraints on the DC inductor~\cite{SysDCB2020Mudar}. Furthermore, the critical converter, power flow, and fault type and location are identified at this stage using the methods presented earlier in Section~\ref{subsec:critical cases}.

\subsubsection{Part B -- Iterative analytical and EMT design}

This part of the method iteratively approximates the DC inductor size based on analytical calculations and EMT simulations to satisfy both the converter and DCCB requirements. In~(\ref{eq:L_con}), all parameters are known, while $\Delta I_{in}$ and $\Delta I_{cab}$ are obtained iteratively from EMT simulations to approximate the inductor value.  Initially, assuming $\Delta I_{in} = 0$ and $\Delta I_{cab} = 0$, the preliminary value of $L_{dc}^{con}$ can be estimated. If the DCCB requirement is more restrictive ($L_{dc}^{con} < L_{dc}^{cb}$), there is no need to proceed with the EMT simulation in Part~B. Otherwise, since~(\ref{eq:L_con}) provides a conservative estimate (due to $\Delta I_{in} = 0$ and the absence of connection cable discharge current between the converter and the fault location), an iterative simulation approach is used. 

By updating $\Delta I_{in}$ and $\Delta I_{cab}$, which depend on the grid configuration (and may be zero), the new value of $L_{dc}^{con}$ is recalculated using~(\ref{eq:L_con}). Part~B is terminated when \eqref{eq:telorance} is met, where $\varepsilon$ is the tolerance between two successive values (e.g., 5\%) and `i' indicates the iteration number. After some iterations, Part~B converges to an $L_{dc}$ value that satisfies~(\ref{eq:L_con}) and~(\ref{eq:Lcb}). However, the designed inductor in Part~B may still be conservative (i.e., its value can be further reduced), as several simplifications were made in deriving~(\ref{eq:L_con}) and~(\ref{eq:Lcb}). Thus, further refinement of the inductor value may be in Part~C.
\begin{equation} \label{eq:telorance}
\frac{|L_{dc(i)}^{ {con}} - L_{dc(i-1)}^{ {con}}| }{L_{dc(i-1)}^{ {con}}}< \varepsilon\, 
\end{equation}
In this design method, any converter model can be used. Consequently, it provides a promising approach when black-box models and industrial vendor models with unknown control implementations are utilized. However, it is important to note that certain converter parameters, such as $S$, $i_{ac}^{max}$, and $K$, must be known to apply the proposed method.

\begin{figure}[!t]
    \centering
    \includegraphics[clip, trim=0cm 0.7cm 0cm 1.1cm, width=0.5\linewidth]{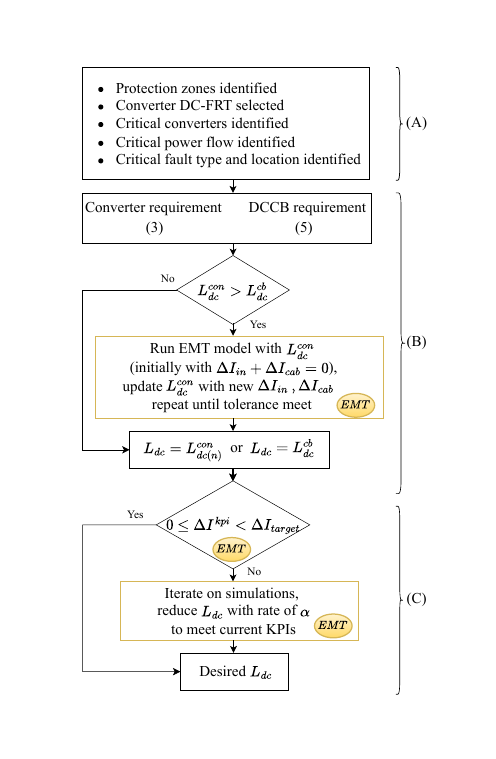}
    
   \caption{Core component sizing algorithm for a single inductor}
   \label{fig:core-algorithm}
\end{figure}

\subsubsection{Part C -- Final refinement and KPIs}
Upon deriving the analytical method, several simplifying assumptions were made in \eqref{eq:L_con} and \eqref{eq:Lcb}. The converter internal voltage is assumed as a constant-voltage source with an inductor, and a lumped inductor is used for modeling the connection cable as shown in Fig.~\ref{fig:sys_for_derive_Lcon}. In \eqref{eq:Lcb}, the voltage across the inductor is assumed constant and equal to the nominal system voltage. These assumptions may lead to an overestimation of  $L_{dc}$. Since the proposed method uses the EMT models with a detailed representation of these components, the design parameter is iteratively fine-tuned to meet the design requirements.

\begin{figure}
    \centering
    \includegraphics[clip, trim=1cm 0.6cm 0.5cm 0.5cm, width=0.4\linewidth]{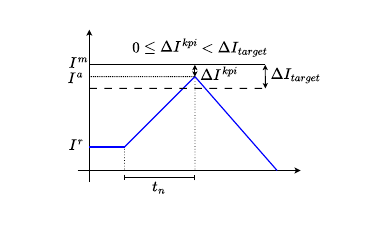}
    
   \caption{Current margin criterion for inductor refinement}
   \label{fig:margin}
\end{figure}

To quantify whether the converter and DCCB requirements are met while the $L_{dc}$ is minimal, a current margin KPI is used as shown in Fig.~\ref{fig:margin}. It is defined as the difference between the maximum current limit and the current achieved with the designed inductor. For the converter, this is based on the arm current, as shown in~(\ref{eq:arm_current}), and for the DCCB, it is based on the DCCB current, as shown in~(\ref{eq:cb_current}).
\begin{align} 
    \Delta I_{con}^{kpi} &= \frac{K i_{a}^{r} - i_{a}^{a}}{i_{a}^{r}(K-1)}
     \label{eq:arm_current} \\
    \Delta I^{kpi}_{cb\,\,\,} &= \frac{I_{cb}^{max} - I_{cb}^a}{I_{cb}^{ max} - I_{cb}^r}
     \label{eq:cb_current}
\end{align}

Based on \eqref{eq:arm_current},  $i_{a}^{a}$ is the achieved peak arm current across all six arm currents when the sized $L_{dc}$ is used.  In \eqref{eq:cb_current}, $I_{cb}^{\max}$ is the DCCB current limit, $I_{cb}^{r}$ is the rated DCCB current, and $I_{cb}^{a}$ is the achieved DCCB current when using the sized inductor. 

The final criterion in Part~C is the minimum of (\ref{eq:arm_current}) and (\ref{eq:cb_current}), which identifies the limiting component as defined in \eqref{eq:minKPI}. A value of $\Delta I^{kpi} = 0$ indicates that the current has reached its maximum allowable limit. Therefore, the current KPI should satisfy the condition $0~\leq~\Delta I^{kpi}~<~\Delta I_{\mathrm{target}}$. As shown in Fig.~\ref{fig:margin}, $\Delta I_{\mathrm{target}}$ is a target that the algorithm aims to meet by reducing the DC inductor size. Additionally, it can also serve as a design margin. In each iteration, the new value is obtained using $L_{dc}^{new} = L_{dc} (1-\alpha)$, where $0<\alpha<1$ is the reduction rate. To achieve faster convergence in Part~C, a dynamic reduction rate is used instead of a fixed one. A large value of $\alpha$ increases the convergence rate, but it also increases the risk of a large step reduction that may violate the KPIs. Thus, a safe value for that is recommended as $\alpha~\leq~40\% \,\Delta I^{kpi}$. 

\begin{equation} \label{eq:minKPI}
    \Delta I^{kpi} = \min\!\left(\Delta I^{kpi}_{ {con}},\; \Delta I^{kpi}_{ {cb}}\right)
\end{equation}
\subsection{Generalized component sizing algorithm}

\begin{figure}
    \centering
    \includegraphics[clip, trim=0cm 0.5cm 0cm 0.4cm, width=0.5\linewidth]{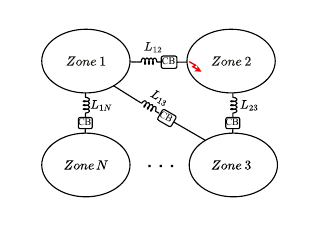}
      
     \caption{Conceptual HVDC system with N protection zones}
     \label{fig:General}
\end{figure}

\begin{figure}
    \centering
    \includegraphics[clip, trim=0cm 0.8cm 0cm 1.0cm, width=0.5 \linewidth]{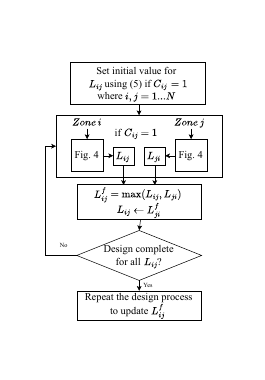}
   
  \caption{Generalized inductor design method for protection systems with multiple protection zones}
  \label{fig:General_Algorithm}
\end{figure}

To demonstrate an effective design for a generic system with multiple protection zones, a conceptual system with $N$ protection zones is considered (Fig.~\ref{fig:General}), where neighboring zones are separated by a DCCB and an inductor. A generalized method for protection component design that accounts for the requirements of multiple protection zones is presented in Fig.~\ref{fig:General_Algorithm}. Initially, inductors in the system are set to values that satisfy the series DCCB requirements based on~(\ref{eq:Lcb}). In the next stage, for two proxy protection zones (directly connected via a DCCB), the inductor is designed using the core algorithm shown in Fig.~\ref{fig:core-algorithm}. The converter requirements apply only to zones that contain converters. For instance, considering Fig.~\ref{fig:General}, to design the inductor $L_{12}$, a fault is applied in Zone~2, and the inductor is designed based on Fig.~\ref{fig:core-algorithm}. Here, $C_{ij} = 1$ means protection zone~i is connected to zone~j with a DCCB. Otherwise $C_{ij} = 0$. The same process is applied for Zone~2 to determine $L_{21}$. Therefore, the final inductor value is set to the maximum of two values.

After designing the first DC inductor between two zones, the newly designed inductor is updated and replaced with its initial value in the system, and the process is repeated for the remaining inductors until all are designed. Additionally, the initial values of nearby inductors impact the target inductor value. To account for this interdependency during design so that the initial value of the nearby inductor has less impact on the designed inductor, an additional iteration is included in the final stage of the algorithm (Fig.~\ref{fig:General_Algorithm}) for the final inductor design. In this iteration, the previously designed values are used as initial inductor values. 


%% file: 3-Case_Studies.tex
\section{Case Studies}

\subsection{Case study description}

\begin{figure}[!t]
    \centering
    \includegraphics[clip, trim=0.9cm 1.2cm 0.8cm 1.1cm, width=0.5\linewidth]{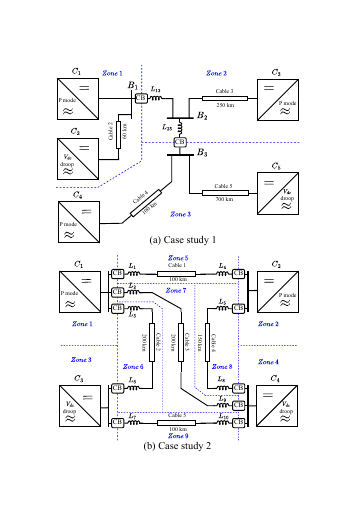}
      \caption{Schematic per-pole diagram of considered case studies: (a) Case Study~1 partially selective protection strategy, (b) Case Study~2 fully selective protection strategy}
    \label{fig:Cases}
\end{figure}

To demonstrate the application of the proposed method, two case studies are presented in this paper, as shown in Fig.~\ref{fig:Cases}. In Case Study~1, a five-terminal bipolar system with a partially selective protection strategy is used. This system is divided into three protection zones by two DCCBs in series with DC inductors. Conversely, in Case Study~2, a four-terminal bipolar system with a fully selective protection strategy is evaluated. In both case studies, it is assumed that a suitable system-level design has already been carried out, providing DCCB locations as inputs to this component design stage. The half-bridge modular multilevel converter (HB-MMC), currently the preferred topology in industry~\cite{Cwikowski-2018}, is employed in this study, with all converters sharing the same parameters and technology. Detailed system and converter data are provided in Table~\ref{tab:grid_parameters}.  

The key design parameters are $K$, $I_{cb}^{\max}$, $t_{cb}$, and $L_{dc}$. Table~\ref{tab:DCCB_params} presents different possible values for the key design input parameters (resulting in different design scenarios), considering some viable parameters for industrial DCCBs~\cite{DCCB2020China,scibreak2020} and converter overcurrent ratings~\cite{Patrick2023,Cwikowski-2018}.

\begin{table}
    \centering
    \caption{System and component parameters }
    \label{tab:grid_parameters}
    \begin{tabular}{l c c c}
    \hline
         Parameter                           & Symbol            & Value & Unit \\
    \hline
         Rated apparent power (per pole)     & $S$               & 1034   & MVA \\
         Rated active power (per pole)       & $P$               & 1000   & MW \\
         Rated DC voltage                    & $U_{dc}$          & 525    & kV \\
         Rated AC voltage (converter side)   & $u_{ac}$          & 273    & kV (rms) \\
         Rated peak arm current              & $i_{a}^{r}$       & 2.2    & kA \\
         Converter AC current limit          & $i_{ac}^{max}$    & 1.2    & pu \\
         Converter overcurrent rating        & $K$               & 1.5--2 & pu \\
         Converter arm resistance            & $R_{arm}$         & 0.4    & $\Omega$ \\
         Converter arm inductor              & $L_{arm}$         & 15     & mH \\
         Submodule capacitance               & $C_{sm}$          & 6000   & $\mu$F \\
         Number of SM per arm                & $N$               & 276    & -- \\
         DCCB current rating                 & $I_{cb}^{r}$      & 3      & kA \\ [1pt]
    \hline  
    \end{tabular}
\end{table}
\begin{table}
    \centering
    \caption{Considered key input design parameters}
    \label{tab:DCCB_params}
    \begin{tabular}{c|cc}
        \hline
        \text{Converter} & \multicolumn{2}{c}{\text{DCCB}} \\
        \hline
        $K$ & $I_{cb}^{\max}$ (kA) & $t_{cb}$(ms) \\ [1pt]
        \hline
        1.5, 2 & 12, 24 & 2, 3, 5 \\
        \hline
    \end{tabular}
\end{table}
\subsection{Component models}
 The systems are modeled in the EMT environment (PSCAD). The cables are model by a frequency-dependent (phase) model for 525 kV. The converters utilize a Type~5 (average) model~\cite{TB604} in these example cases, though vendor models could be used for an industrial study. Additionally, a typical control scheme is implemented for the converters, including outer-level controls (such as DC voltage droop and active power controllers) and inner-level controls (such as a circulating current suppression controller with typical parameters)~\cite{cigre_tb832}. Furthermore, internal overcurrent protection is incorporated into the converters~\cite{Patrick2023,Cwikowski-2018}. For the DCCBs, a Type~5 model~\cite{Cigre-TB873} is adopted, utilizing generic data for the metal-oxide varistors (MOVs)~\cite{Bucher2014}. The AC grids are represented by Thevenin equivalents with specified short-circuit ratios. The simulation time step is 20~$\mu\mathrm{s}$.

%% file: 4-Results.tex
\section{Results of Component Sizing for Case Studies}

\subsection{{Case study 1: partially selective protection strategy}}

Component design is performed for Case Study~1, shown in Fig.~\ref{fig:Cases}~(a), using the generalized algorithm shown in Fig.~\ref{fig:General_Algorithm}. In this Case Study, the grid is divided by the protection system into three protection zones~\cite{van2025shortlisting}, each separated by a DCCB in series with a DC inductor. Regarding converter DC-FRT requirements, it is assumed that converters in the healthy zone must remain active and not block during an out-of-zone fault. 

Since all converters have identical parameters (Table~\ref{tab:grid_parameters}), the converter directly connected to the DCCB, or the one connected through the shortest line, is considered critical in each zone. Thus, $C_1$ in Zone~1, $C_3$ in Zone~2, and $C_4$ in Zone~3 are the critical converters. In general, \eqref{eq:L_con} and \eqref{eq:I_con_max} can be used to identify the critical converter when the converters have different parameters.
The critical power flows that result in maximum loading for the critical converters and DCCBs are identified in Table~\ref{tab:critical-PFs}. A positive converter active power setpoint indicates rectifier mode. In the example considered, converter $C_1$ connects to an offshore wind farm, so its minimum active power setpoint is zero. Additionally, it is assumed that converters $C_2$ and $C_5$ operate in DC-voltage droop control mode. The critical fault type is considered to be a pole-to-ground (P--g) fault with zero fault resistance~\cite{promotion2017d42}. Although the traveling-wave effect~\cite{cwikowski2016} can cause a fast rise in fault current during a short time window for non-terminal faults (Fig.~\ref{fig:CFL}), a terminal fault causes the highest fault current given the operation time of the chosen DCCBs (at least 2~ms). Thus, the critical fault location in this Case Study is at the terminal of the adjacent cable on the other zone, resulting in $\bar{U}_{m}=0$ in \eqref{eq:L_con}.

\begin{table}[!t]
\small
\caption{Critical converter setpoints for each protection zone and their corresponding inductor design (Case Study~1)}
\label{tab:critical-PFs}
\centering
\begin{tabular}{ccccccc}
\hline
\multicolumn{2}{c}{ } & \multicolumn{5}{c}{Converter active power setpoint (pu)} \\
% \cline{1-7}
Zone & Inductor & C1 & C2 & C3 & C4 & C5 \\
\hline
$1$ & $L_{12}$ & $+1$  & $-$ & $-1$  & $-1$  & $-$ \\
$2$ & $L_{21}$ & $~~0$  & $-$ & $+1$  & $+1$  & $-$ \\
$2$ & $L_{23}$ & $+1$  & $-$ & $+1$  & $-1$  & $-$ \\
$3$ & $L_{32}$ & $~~0$  & $-$ & $-1$  & $+1$  & $-$ \\
\hline
\end{tabular}
\end{table}

\subsubsection{Inductor sizing to meet Zone~1 requirements}
 
\begin{table}[!t]
    \centering
    \caption{Inductor design results for Zone~1 requirements considering different input parameters (Case Study~1)}
    \label{tab:results1}
    % \footnotesize
    % \resizebox{\columnwidth}{!}{%
    \begin{tabular}{c ccc c cc}
    \hline
    & \multicolumn{3}{c}{Input parameters} 
    & \multicolumn{1}{c}{Zone~1} 
    & \multicolumn{2}{c}{KPIs} \\
    \cline{2-4} \cline{6-7}
    Scen.\rule{0pt}{8pt}
    & $K$ (pu) & $I_{cb}^{\max}$ (kA) & $t_{cb}$ (ms) 
    & $L_{12}$ (mH) 
    & $\Delta I_{con}$ (\%) & $\Delta I_{cb}$ (\%) \\[1pt]
    \hline
    1& 1.5 & 12 & 2 & 165 & $\mathbf{5}$  & 31 \\
    2& 1.5 & 12 & 3 & 235 & $\mathbf{4}$ & 30 \\
    3& 1.5 & 12 & 5 & 414 & $\mathbf{3}$  & 33 \\
    % \hdashline[1pt/2pt]
    4& 2   & 12 & 2 & 109 & 32& $\mathbf{2}$  \\
    5& 2   & 12 & 3 & 161 & 30& $\mathbf{4}$  \\
    6& 2   & 12 & 5 & 267 & 5 & $\mathbf{4}$  \\
    \hline
    \end{tabular}
    % }
\end{table}

\begin{figure}
    \centering
    \includegraphics[clip, trim=0.8cm 0.5cm 0.2cm 0.8cm,width=.6\linewidth]{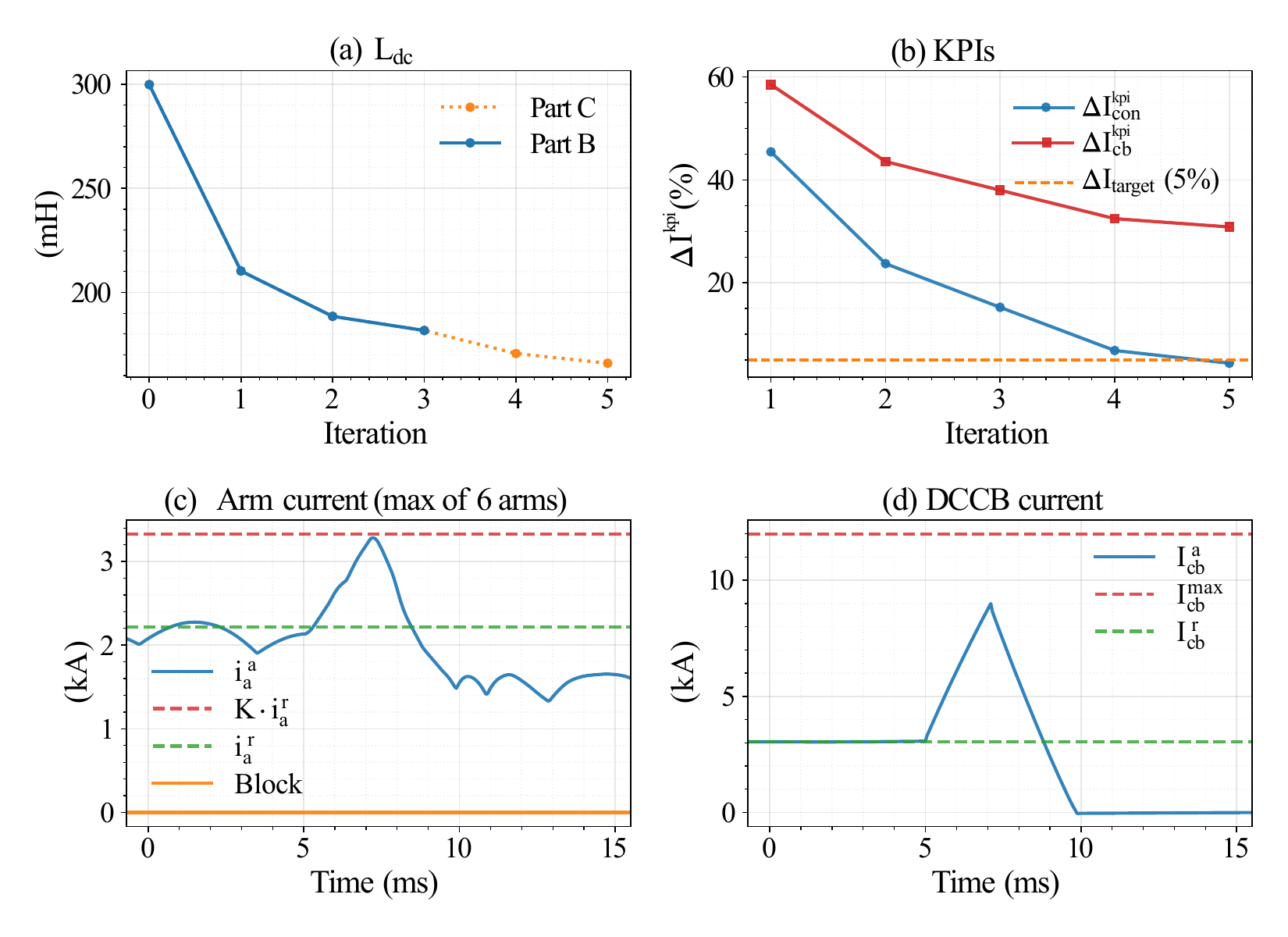}
    \caption{Converter as a limiting factor, evaluated at $L_{12}=165$~$\mathrm{mH}$ (Scenario~1 of Case Study~1)} 
    \label{fig:Case1}

\end{figure}
To illustrate the application of the core algorithm in Fig.~\ref{fig:core-algorithm}, it is first applied to Zone~1 to size $L_{12}$ to satisfy the protection requirements of this zone. Accordingly, the corresponding critical power setpoints from the first row of Table~\ref{tab:critical-PFs} are selected. For a P--g fault at the beginning of cable~3 (bus~$B_2$), the algorithm is executed for some input parameter combinations given in Table~\ref{tab:DCCB_params}, and the resulting values of $L_{12}$ are summarized in Table~\ref{tab:results1}. According to the results, the DC inductor size varies depending on the input sets. Scenario~3 with $L_{12}=414\,\mathrm{mH}$ and Scenario~4 with $L_{12} = 109\,\mathrm{mH}$ require the largest and smallest inductor, respectively. Based on the current margin KPIs, it is evident that the method can size the inductor to achieve the target threshold margin ($\Delta I_{target} = 5\%$). Considering the competent constraint, depending on the input parameters, the converter arm current is a limiting factor in scenarios 1--3 ($\Delta I^{kpi}_{con} < \Delta I^{kpi}_{cb}$), while in Scenarios~4--6, the DCCB constraint becomes a limiting factor.

In Fig.~\ref{fig:Case1}, the results of Scenario~1 are presented in detail. Plot~(a) shows that Part~B of the algorithm converges to approximately $L_{dc}=182$~mH after three iterations. However, due to simplifications in the analytical formulation, further refinement is carried out in Part~C, where its final minimum feasible value of $L_{dc}=165$~mH. As shown in plot~(b), the algorithm terminates after five iterations, when the converter current margin meets the target threshold. Plots~(c) and~(d) show the converter arm current and DCCB current, respectively, evaluated with the final designed inductor $L_{dc} = 165\,\mathrm{mH}$, demonstrating that the converter arm current approaches its limit first. 

\begin{figure}
    \centering
    \includegraphics[clip, trim=0.8cm 0.5cm 0.2cm 0.8cm,width=.6\linewidth]{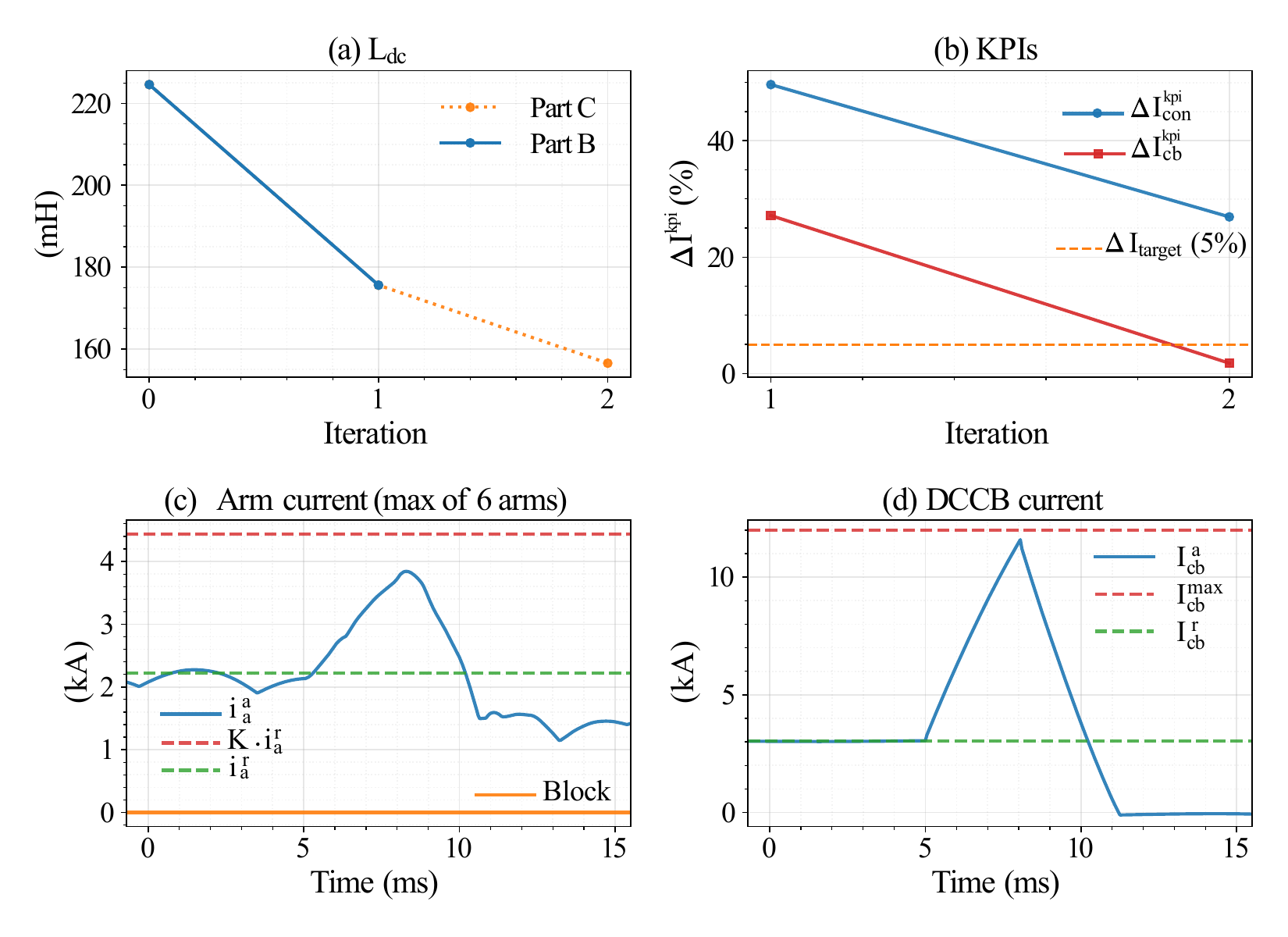}
    \caption{DCCB as a limiting factor, evaluated at $L_{12}=161$~$\mathrm{mH}$ (Scenario~4 of Case Study~1)} 
    \label{fig:Case8}
\end{figure}

 On the other hand, Fig.~\ref{fig:Case8} presents the result for Scenario~4, where the DCCB constraint is the limiting factor ($\Delta I^{kpi}_{cb} < \Delta I^{kpi}_{con}$).  Comparing the results of Scenario~1 and Scenario~4, the former converges to its final value in five iterations, whereas the latter converges in only two iterations.

\subsubsection{Inductor sizing to meet all zones requirements}

As shown in Fig.~\ref{fig:Cases}(a), the system consists of three zones and two inductors. Each inductor must meet the requirements of both connected zones. Specifically, $L_{12}$ is determined by considering the requirements of both Zone~1 (associated with $L_{12}$) and Zone~2 (associated with $L_{21}$). Similarly, $L_{23}$ is designed based on the requirements of Zone~2 and Zone~3 (associated with $L_{32}$). In total, the core algorithm (Fig.~\ref{fig:core-algorithm}) should be executed four times per input set. 
The results of inductor sizing for all considered input parameters in Table~\ref{tab:DCCB_params} with respect to all protection zones are provided in Table~\ref{tab:All_zones_results}. The final values for the two inductors in the system are in the last columns. 

\begin{table}
\centering
\caption{Final inductor design results for Case Study~1}
\label{tab:All_zones_results}
% \footnotesize
% \resizebox{\columnwidth}{!}{
\begin{tabular}{c ccc cccc cc}
\hline
& \multicolumn{3}{c}{Input parameters}
& \multicolumn{4}{c}{Zone requirements (mH)}
& \multicolumn{2}{c}{Final (mH)} \\
\cline{2-4} \cline{5-8} \cline{9-10}
Scen.\rule{0pt}{8pt}
& $K$ & $I_{cb}^{\max}$ (kA) & $t_{cb}$ (ms)
& $L_{12}$ & $L_{21}$ & $L_{23}$ & $L_{32}$
& $L_{12}^f$ & $L_{23}^f$ \\[1pt]
\hline
1  & 1.5 & 12 & 2 & {  \textbf{165}} & {  101} & { \textbf{105}} & { 98}  & \textbf{{  165}} & \textbf{{ 105}} \\
2  & 1.5 & 12 & 3 & {  \textbf{235}} & {  144} & { 156} & { \textbf{164}} & \textbf{{  235}} & \textbf{{ 164}} \\
3  & 1.5 & 12 & 5 & {  \textbf{413}} & {  306} & { \textbf{404}} & { 283} & \textbf{{  413}} & \textbf{{ 404}} \\
% \hdashline[1pt/2pt]
4  & 1.5 & 24 & 2 & {  \textbf{165}} & {  70}  & { 87}  & { \textbf{99}}  & \textbf{{  165}} & \textbf{{ 99}}  \\
5  & 1.5 & 24 & 3 & {  \textbf{236}} & {  111} & { 157} & { \textbf{164}} & \textbf{{  236}} & \textbf{{ 164}} \\
6  & 1.5 & 24 & 5 & {  \textbf{414}} & {  218} & { \textbf{404}} & { 283} & \textbf{{  414}} & \textbf{{ 404}} \\
% \hdashline[1pt/2pt]
7  & 2   & 12 & 2 & {  \textbf{109}} & {  92}  & { 82}  & { \textbf{94}}  & \textbf{{  109}} & \textbf{{ 94}}  \\
8  & 2   & 12 & 3 & {  \textbf{161}} & {  128} & { 114} & { \textbf{132}} & \textbf{{  161}} & \textbf{{ 132}} \\
9  & 2   & 12 & 5 & {  \textbf{267}} & {  200} & { 177} & { \textbf{215}} & \textbf{{  267}} & \textbf{{ 215}} \\
% \hdashline[1pt/2pt]
10 & 2   & 24 & 2 & {  \textbf{80}}  & {  37}  & { 43}  & { \textbf{49}}  & \textbf{{  80}}  & \textbf{{ 49}}  \\
11 & 2   & 24 & 3 & {  \textbf{122}} & {  64}  & { 78}  & { \textbf{96}}  & \textbf{{  122}} & \textbf{{ 96}}  \\
12 & 2   & 24 & 5 & {  \textbf{254}} & {  127} & { 168} & { \textbf{176}} & \textbf{{  254}} & \textbf{{ 176}} \\
\hline
\end{tabular}
% }
\end{table}

According to the results in Table~\ref{tab:All_zones_results}, considering the final inductor value $L_{12}^f$, in all considered input parameters, Zone~1 requirement is the limiting one, as it requires a larger value of inductor ($L_{12}$) compared with the $L_{21}$. The reason is that the critical converter in Zone~1 ($C_1$) is directly connected to DCCB, whereas the critical converter in Zone~2 ($C_3$) is connected to DCCB through a 250~km cable. The connection cable reduces the required inductor value, as indicated in (\ref{eq:L_con}). In addition, due to the propagation delay of the cable, the contribution of $C_3$ to a fault in Zone~1 is delayed. This delayed response further alleviates the inductor requirement.
For the second inductor ($L_{23}^f$), looking at the values for $L_{23}$ and $L_{32}$, it is clear that the critical zone determining the final inductor value can vary depending on the input parameters. This highlights the need for a design based on all protection zones, as it is not always clear which is the primary limiting zone. Furthermore, when the number of nearby inductors is no more than one, such as Case Study~1, the effect of interdependency between the inductors is low, and therefore, there is no need for the final step in~Fig.\ref{fig:General_Algorithm}.

\begin{figure}
    \centering
    \includegraphics[clip, trim=0.3cm 0.2cm 0.2cm 0.3cm,width=0.6\linewidth]{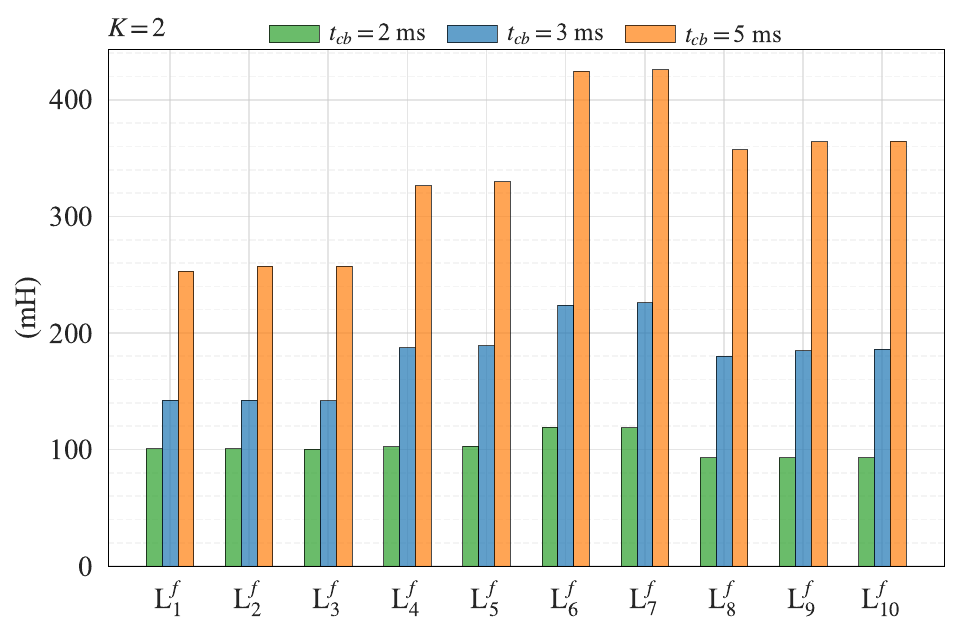}
    \caption{Final inductor design results for Case Study~2 based on Fig.\ref{fig:General_Algorithm}}
    \label{fig:L1-10}
\end{figure}

\begin{figure}
    \centering
    \includegraphics[clip, trim=0.6cm 0.6cm 1cm 0.3cm,width=0.6\linewidth]{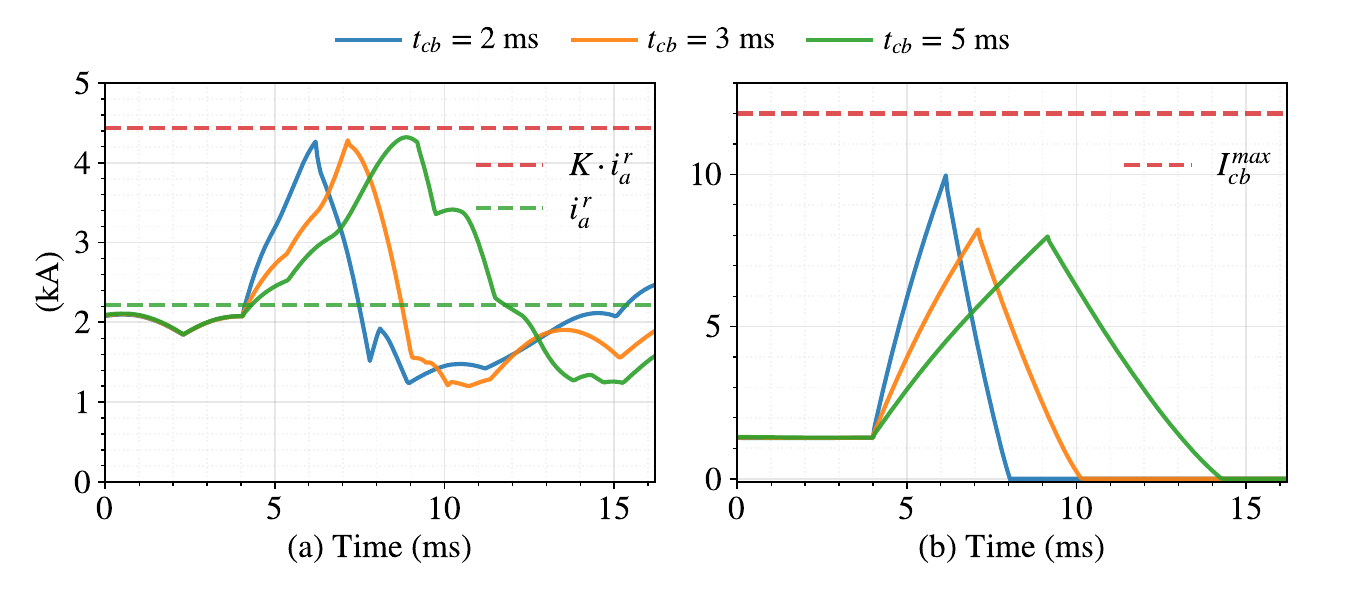}
    \caption{Impact of various scenarios on (a) converter $C_2$ arms current and (b) DCCB current connection with $L_5$, for a fault at beginning of cable~4 from $C_2$ (Case study~2)}
    \label{fig:L5}
\end{figure}

\begin{figure}[t]
    \centering
    \includegraphics[clip, trim=0.6cm 0.6cm 1cm 0.7cm,width=0.6\linewidth]{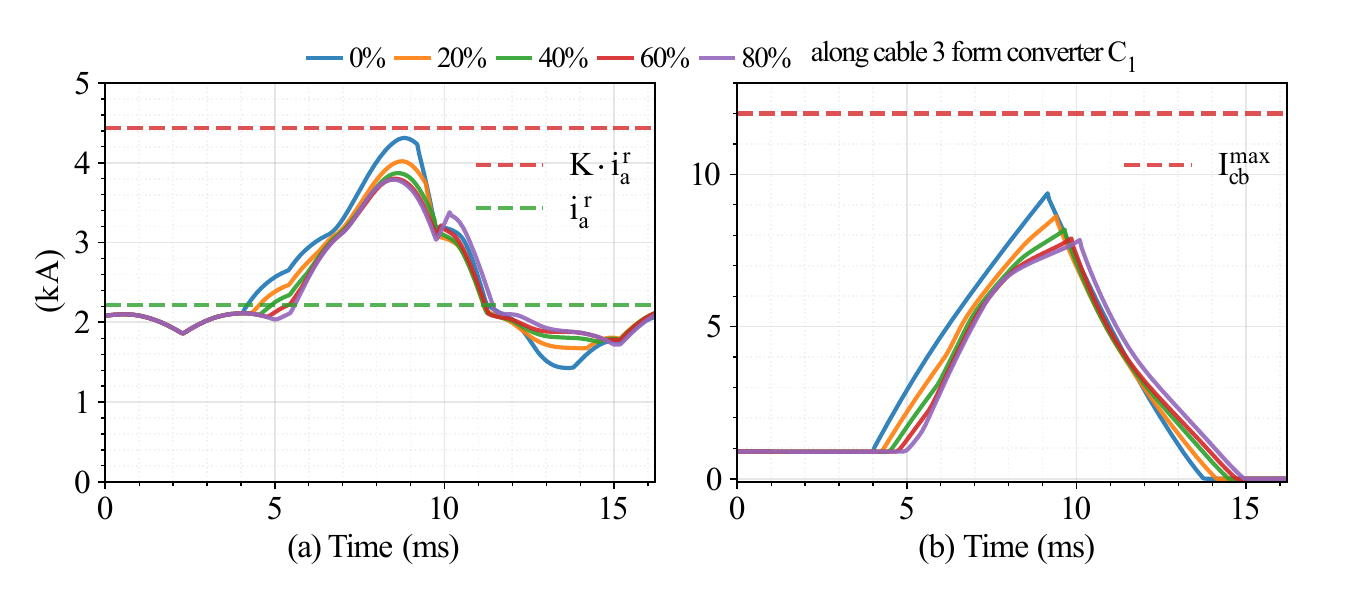}
    
    \caption{Impact of fault location on (a) converter $C_1$ arms current and (b) DCCB current for $K=2$, $t_{cb}=5$ ms, $I_{cb}^{max}=12$ kA with different fault location along the cable~3 from $C_1$ (Case study~2)}
    \label{fig:FL}
\end{figure}

\subsection{Case study 2: fully selective protection strategy}

To evaluate the effectiveness of the proposed method for a meshed system with a fully selective protection strategy, Case Study~2 is considered, as shown in Fig.~\ref{fig:Cases}(b). In this case, there are ten DCCBs, each in series with a DC inductor and nine protection zones. Regarding DC-FRT requirements, none of the converters are allowed to block during out-of-zone faults. All four converters are critical to their respective zones. With regard to critical power flows, to design inductors $L_1$ to $L_5$, converters $C_1$ and $C_2$ should operate in rectifier mode at rated power, while for $L_6$ to $L_{10}$, converters $C_3$ and $C_4$ should be in rectifier mode at rated power. This means that, in total, two sets of power flows should be considered. 

By applying the proposed algorithm outlined in Fig.~\ref{fig:General_Algorithm}, the designed inductor values are shown in Fig.~\ref{fig:L1-10}. In this Case Study, the number of scenarios is reduced to three: $I_{cb}^{\max} = 12~\text{kA}$ and $K=2$ are kept constant, while three values of $t_{cb}$ are considered. Furthermore, in each scenario, all DCCBs in the system are assumed to be identical. 

Fig.~\ref{fig:L5} shows the arm current of converter $C_2$ and the current of the DCCB in series with inductor $L_5$ for the three scenarios. This demonstrates that when a fault occurs in cable~4, the designed inductor sufficiently limits the rate of rise of the fault current, ensuring that both the converter and DCCB requirements are met. Additionally, Fig.~\ref{fig:FL} investigates the impact of the fault location, where a P--g fault is applied along cable~3, and the arm current of converter $C_1$, along with the current of the DCCB in series with $L_2$, are monitored. These results correspond to the $t_{cb}=5$~ms scenario, which imposes the most demanding conditions on the inductor. The results confirm that a terminal fault produces the highest fault current, and the designed inductor correctly limits it below the converter arm limits.

\begin{figure}
    \centering
    \includegraphics[clip, trim=0.2cm 0.7cm 0cm 0.6cm,width=0.6\linewidth]{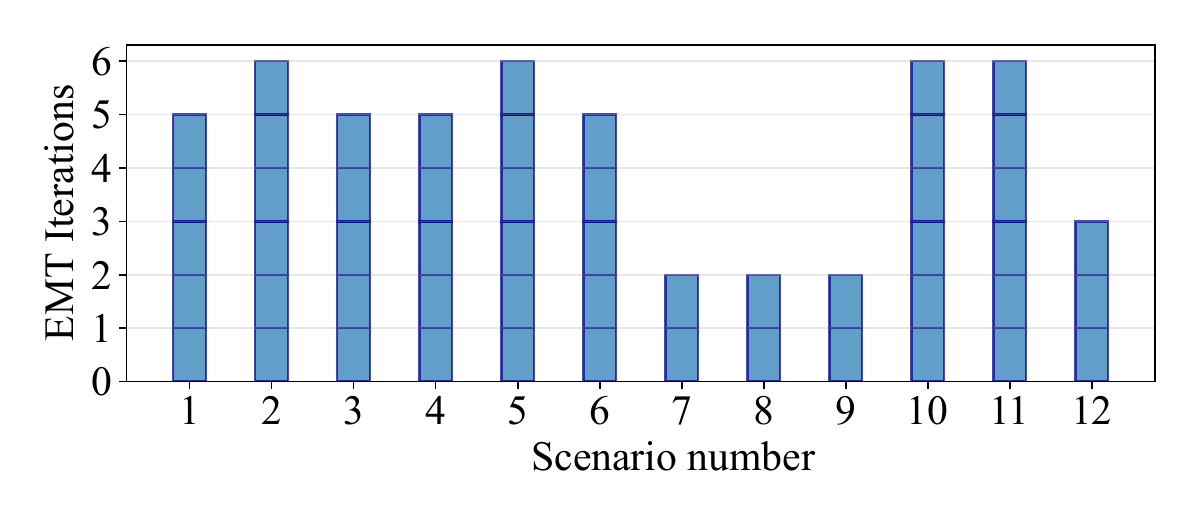}
\caption{Number of simulations required for sizing the component of a single protection zone}
    \label{fig:iteration}
\end{figure}

\subsubsection{Impact of system configuration and nearby inductors} 
As shown in \eqref{eq:L_con}, the infeed current from adjacent cables to the converter affects the inductor value. A higher number of adjacent cables increases the infeed current, thereby reducing the required inductance. For example, according to Fig.~\ref {fig:L1-10}, $L_1$ has a smaller value than $L_4$, which can be explained by the impact of two adjacent cables, while $L_4$ is influenced by only one adjacent cable (for $t_{cb}=3, 5$ ms). 

Additionally, there is an interdependency between nearby inductors. The higher the inductance of a nearby inductor, the more it limits the infeed current during a fault and requires a higher inductance. However, inductors that are far away have less impact on the local inductor value. This interdependency is addressed in the proposed algorithm in Fig.\ref{fig:General_Algorithm}. 

\subsubsection{Impact of converter control mode} 
A comparison between $L_4$ and $L_6$ in Fig.~\ref{fig:L1-10} reveals that they share almost identical conditions regarding the number of adjacent cables and converter parameters, except for the converter control mode. Converter $C_3$ operates under DC voltage (droop) control ($V_{dc}$), whereas converter $C_2$ operates in PQ control mode. The results for this Case Study indicate that voltage control requires a larger inductor value than active power control. However, it is worth noting that different control implementations and parameters may yield different results. Therefore, further investigation into this aspect is required in future work.

\subsection{Efficiency of the method} 

By utilizing the hybrid method, the proposed approach significantly reduces the number of required simulations compared to simulation-based methods~\cite{DCR-ACDC2026,hart2025impact,Ilka2021}, where the number of estimated required iterations is in the order of hundreds to thousands, and it can take a few days to finish~\cite{Ilka2021}, depending on the case. As shown in Fig.~\ref{fig:iteration} for a single DC inductor, the number of EMT iterations varies across scenarios, reaching a maximum of six. The average number of simulation iterations per scenario for Cases Sduies~1 and Cases Sduies~2 is 9 and 88, respectively, which can take a few minutes to an hour to run.  
As shown in Fig.~\ref{fig:General_Algorithm}, the number of executions of the core algorithm (Fig.~\ref{fig:core-algorithm}) and the total number of simulation iterations depend on the number of DC inductors and critical converters in the system.

Furthermore, identifying the critical simulation cases for component-level design can significantly reduce the number of simulations required and improve the design process's efficiency. For example, if a single critical fault location is identified and used as the design basis, it eliminates the need to evaluate multiple alternative fault locations. This targeted approach simplifies the analysis while ensuring that the design is based on the critical conditions.

%% file: 5-Conclusion.tex
\section{Conclusion} 

A hybrid analytical-EMT method was proposed for systematically specifying protection component parameters in HVDC grids. The method ensures that the protection system requirements and component constraints are satisfied. It can be used for both partially and fully selective protection strategies. Two measures were used to enhance the efficiency of the method. First, critical cases were identified to reduce the number of design scenarios, thereby improving the efficiency of the design process. Second, a hybrid analytical–EMT approach was adopted, in which the analytical part initializes and guides the EMT simulations, resulting in faster convergence and fewer simulations -- orders of magnitude faster for the considered case studies. The proposed method provides high-accuracy results by directly incorporating detailed component models. In particular, the possibility of using vendor models ensures high-fidelity results. Depending on the system-level and component-level requirements and constraints, such as DCCB technology, the required DC inductor can vary. In the considered case studies, the inductance varied between around $50$ and $450$~mH.